\mag=\magstep1

\documentclass[12pt]{article}

\usepackage{latexsym}
\usepackage{amssymb}
\usepackage{amsmath}

\newcommand{\niklabel}[1]{\label{#1}}

\def\num#1{\par\medskip\noindent{#1}}

\def\{{\lbrace}
\def\}{\rbrace}

\def\cl{{\cal C}\!\ell}

\def\R{{\mathbb R}}
\def\C{{\mathbb C}}

\def\U{{\rm U}}
\def\SU{{\rm SU}}

\def\diag{{\rm diag}}
\def\Even{{\rm Even}}
\def\Odd{{\rm Odd}}

\def\Sp{{\rm Sp}}

\def\sp{{\rm sp}}

\def\be{\begin{equation}}
\def\ee{\end{equation}}

\def\Mat{{\rm Mat}}

\def\T{{\rm T}}

\def\be{\begin{equation}}
\def\ee{\end{equation}}

\usepackage{makeidx}
\makeindex

\begin{document}

\title{Mass generation mechanism for spin 1/2 fermions in Dirac--Yang--Mills model
equations with a symplectic gauge symmetry}

\author{Nikolay Marchuk}

\maketitle

\begin{abstract}
In the Standard Model of electroweak interactions the fundamental
fermions acquire masses by the Yukawa interaction with the (spin
0) Higgs field. In our model spin 1/2 fermions acquire masses by
an interaction with (spin 1) gauge field with symplectic symmetry.
\end{abstract}

In  \cite{RJMP1,RJMP2,TMP,25,mybook} we develop a new approach to
field theory, which based on so called {\em model equations of
field theory}. In this paper we introduce Dirac--Yang--Mills model
equations for spin 1/2 fermions interacting with two gauge fields
simultaneously. One field has a unitary gauge symmetry and another
has a symplectic gauge symmetry. There is no fermion's mass ($m$)
term in the model Dirac equation. But there is the term $3m^3/16$
in the right hand part of the Yang--Mills equations with
symplectic gauge symmetry. Hence, the constant $3m^3/16$ can be
considered as a constant (charge) describing the interaction of a
fermion with the symplectic gauge field.

\medskip

\medskip

\noindent{\bf Clifford algebra}. Let $\cl(1,3)$ be the complex
Clifford algebra \cite{Lounesto} with the unity element $e$ and
with generators $e^a$, $a=0,1,2,3$, which satisfy the relations
$$
e^a e^b + e^b e^a=2\eta^{ab}e,\quad a,b=0,1,2,3,
$$
where $\eta=\|\eta^{ab}\|=\diag(1,-1,-1,-1)$ is the diagonal
matrix.

Let $\cl_k(1,3)$, ($k=0,1,2,3,4$) be subspaces of rank $k$
Clifford algebra elements and $\cl_\Even(1,3)$, $\cl_\Odd(1,3)$ be
the subspaces of even and odd Clifford algebra elements
respectively. By $\cl^\R(1,3)$ denote the real Clifford algebra.

 Denote
$\beta=e^0\in\cl(1,3)$. Consider {\em an operation of
pseudo-Hermitian conjugation}  $* : \cl(1,3) \to \cl(1,3)$ such
that $(e^a)^*=e^a$, $a=0,1,2,3$ and
$$
(\lambda U)^*=\bar\lambda U^*,\quad (UV)^*=V^*U^*,\quad
(U+V)^*=U^*+V^*,
$$
where $U,V$ are arbitrary elements of $\cl(1,3)$ and
$\lambda\in\C$. Now we can define {\em an operation of Hermitian
conjugation} of Clifford algebra elements by the formula
\cite{Marchuk:Shirokov}
$$
U^\dagger=\beta U^*\beta.
$$

\noindent{\bf Symplectic Lie group and its real Lie algebra}.
Consider {\em the real symplectic Lie group} of matrices of even
order $n=2m$ and its Lie algebra
\begin{eqnarray*}
\Sp(m,\R) &=& \{U\in\Mat(n,\R) : U^T S U=S\},\\
\sp(m,\R) &=& \{u\in\Mat(n,\R) : u^T S =-S u\},
\end{eqnarray*}
where $U^T$ is the transposed matrix, $S$ is the block matrix
$$
S=\left( \begin{array}{cc} 0 & -I_m\\ I_m & 0\end{array}\right),
$$
and $I_m$ is the identity matrix of order $m$. Note that
$S^2=-{\bf1}$ (${\bf1}$  is the identity matrix of order $2m$).

\medskip

\noindent{\bf Symplectic Lie group of the Clifford algebra and its
Lie algebra}. Let us define two sets of Clifford algebra elements
\cite{Marchuk:Dyabirov}
\begin{eqnarray*}
\Sp(\cl(1,3)) &=& \{V\in\cl^\R_\Even(1,3)\oplus i\cl^\R_\Odd(1,3) : V^*V=e\},\\
\sp(\cl(1,3)) &=& \{v\in i\cl^\R_1(1,3)\oplus \cl^\R_2(1,3)\}.
\end{eqnarray*}
The set $\Sp(\cl(1,3))$ is closed with respect to the
multiplication of Clifford algebra elements and forms a (Lie)
group. This group is called {\em the symplectic group of Clifford
algebra $\cl(1,3)$}. The set $\sp(\cl(1,3))$ is closed w.r.t. the
commutator $[u,v]=u v-v u$ and forms the Lie algebra.

The following proposition is proved in
{\rm\cite{Marchuk:Dyabirov}}: {\em The group $\Sp(\cl(1,3))$ is
isomorphic to the group $\Sp(2,\R)$ and the Lie algebra
$\sp(\cl(1,3))$ is isomorphic to the Lie algebra $\sp(2,\R)$,
i.e.}
\begin{eqnarray}
\Sp(\cl(1,3)) &\simeq& \Sp(2,\R),\niklabel{sp:sp}\\
\sp(\cl(1,3)) &\simeq& \sp(2,\R).\nonumber
\end{eqnarray}

\medskip

\noindent{\bf Hermitian idempotents}. Let $t\in\cl(1,3)$ be a
nonzero element such that
\begin{equation}
t^2=t,\quad t^\dagger=t, \quad \bar t J=J t,\niklabel{t:cond}
\end{equation}
where $J=-e^1e^3$. Such an element is called {\em a Hermitian
idempotent}.  In particular, we may take the Hermitian idempotents
\begin{eqnarray*}
t_{(1)}&=&\frac{1}{4}(e+e^0)(e+i e^{12}),\\
t_{(2)}&=&\frac{1}{2}(e+e^0),\\
t_{(3)}&=&\frac{1}{4}(3e+e^0+ie^{12}-ie^{012}),\\
t_{(4)}&=&e.
\end{eqnarray*}
A Hermitian idempotent $t$ generates the left ideal $I(t)$, the
two sided ideal $K(t)$, the Lie algebra $L(t)$, and the Lie group
$G(t)$
\begin{eqnarray}
I(t) &=& \{U\in\cl(1,3) : U=U t\},\nonumber\\
K(t) &=& \{U\in I(t) : U=t U\},\niklabel{IKLG}\\
L(t) &=& \{U\in K(t) : U^\dagger =-U\},\nonumber\\
G(t) &=& \{U\in\cl(1,3) : U^\dagger U=e,\,U-e\in K(t)\}\nonumber.
\end{eqnarray}
\medskip

\noindent{\bf The Minkowski space}. Let $\R^{1,3}$ be the
Minkowski space with cartesian coordinates $x^\mu$, where
$\mu=0,1,2,3$ and $\partial_\mu=\partial/\partial x^\mu$ are
partial derivatives. We use Greek indices
$\mu,\nu,\alpha,\beta,\ldots$ (run from 0 to 3) as tensor indices
relative to coordinates $x^\mu$. The Minkowski metric is given by
the diagonal matrix $\eta$. By
 $\T^r_s$ denote the set of tensor fields of type $(r,s)$  (of rank
 $r+s$) in Minkowski space. By  $\T_{[s]}$ denote the set of rank
 $s$ antisymmetric covariant tensor fields. In the sequel we
 consider tensors with values in Lie algebras. For example, if
 $u_\mu$ is a covector with values in a Lie algebra $L(t)$,
then we write $u_\mu\in L(t)\T_1$.

In what follows the generators $e^0,e^1,e^2,e^3$ of Clifford
algebra $\cl(1,3)$ and the fixed Hermitian idempotent $t$ do not
depend on $x$ and they are scalars of the Minkowski space, i.e.
they do not transform under Lorentzian changes of coordinates.

\medskip

\noindent{\bf Model Dirac--Yang--Mills equations}. Consider {\em
the model Dirac--Yang--Mills equations} \cite{TMP,mybook}
\begin{eqnarray}
&&i h^\mu(\partial_\mu\phi+\phi A_\mu-C_\mu\phi)-m\phi=0,\nonumber\\
&&\partial_\mu A_\nu-\partial_\nu A_\mu-[A_\mu,A_\nu]=F_{\mu\nu},
\niklabel{MDYMW1}\\
&&\partial_\mu F^{\mu\nu}-[A_\mu,F^{\mu\nu}] =\phi^\dagger \beta i
h^\nu\phi\nonumber,\\
&&\partial_\mu h^\nu-[C_\mu,h^\nu]=0,\nonumber
\end{eqnarray}
where
\begin{description}
\item[1.] The vector $i h^\mu=i h^\mu(x)\in\sp(\cl(1,3))\T^1$ is
such that
\begin{equation}
 h^\mu h^\nu+h^\nu h^\mu=2\eta^{\mu\nu}e,\quad \mu,\nu=0,1,2,3. \niklabel{h:cond3}
\end{equation} \item[2.] The element $\phi=\phi(x)\in I(t)$ is a
scalar of Minkowski space (it does not transform under Lorentzian
changes of coordinates) ($\phi(x)\to\phi(x(\acute x))$). \item[3.]
$A_\mu=A_\mu(x)\in L(t)\T_1$. \item[4.]
$F_{\mu\nu}=F_{\mu\nu}(x)\in L(t)\T_{[2]}$. \item [5.] The mass
$m$ is a real constant. \item[6.]
$C_\mu=C_\mu(x)\in\sp(\cl(1,3))\T_1$.
\end{description}
We suppose that the idempotent $t$, the constant $m$, and the
generators of Clifford algebra $e^a$ are known and the variables
$h^\mu,\phi,A_\mu,F_{\mu\nu},C_\mu$ are unknown. In this case
equations  (\ref{MDYMW1}) are called {\em model Dirac--Yang--Mills
equations} (with local symplectic symmetry).

From the first equation in (\ref{MDYMW1}), using the identity
$\frac{1}{4}h^\mu h_\mu=e$, we get the equation
$$
i h^\mu(\partial_\mu\phi+\phi A_\mu-B_\mu\phi)=0,
$$
where
$$
B_\mu=C_\mu-\frac{m}{4}i h_\mu\in\sp(\cl(1,3))\T_1.
$$
If we substitute the expression
$$
C_\mu=B_\mu+\frac{m}{4}i h_\mu\in\sp(\cl(1,3))\T_1
$$
into the equalities (the second equality is a consequence of the
first one)
\begin{eqnarray*}
&&\partial_\mu h^\nu-[C_\mu,h^\nu]=0,\\
&&\partial_\mu C_\nu-\partial_\nu C_\mu-[C_\mu,C_\nu]=0,
\end{eqnarray*}
then we get
\begin{eqnarray}
&&\partial_\mu i h^\nu-[B_\mu,i h^\nu]=\frac{m}{4}[i h_\mu,i
h^\nu],
\niklabel{hB:new}\\
&&\partial_\mu B_\nu-\partial_\nu
B_\mu-[B_\mu,B_\nu]=-\left(\frac{m}{4}\right)^2[i h_\mu,i
h_\nu].\nonumber
\end{eqnarray}
From the first equality if follows that
$$
\partial_\mu h^\mu-[B_\mu,h^\mu]=0.
$$
We denote
$$
G_{\mu\nu}=-\left(\frac{m}{4}\right)^2[i h_\mu,i h_\nu].
$$
Using the relations  (\ref{hB:new}) and the relations
$$
\frac{1}{4}h^\mu h_\mu=e,\quad h^\mu h^\nu h_\mu=h_\mu h^\nu
h^\mu=-2 h^\nu,
$$
we see that
$$
\partial_\mu G^{\mu\nu}-[B_\mu,G^{\mu\nu}]=\frac{3}{16}m^3 i h^\nu.
$$

Therefore we have proved that if the variables $\phi,h^\mu$,
$A_\mu,C_\mu,F_{\mu\nu}$ satisfy conditions (\ref{h:cond3}) and
equations (\ref{MDYMW1}), then the variables
$$
\phi,\,h^\mu,\,A_\mu,\,B_\mu=C_\mu-\frac{m}{4}i h_\mu,\,
F_{\mu\nu},\,G_{\mu\nu}=-\left(\frac{m}{4}\right)^2[i h_\mu,i
h_\nu]
$$
satisfy the equations
\begin{eqnarray}
&&i h^\mu(\partial_\mu\phi+\phi A_\mu-B_\mu\phi)=0,\nonumber\\
&&\partial_\mu A_\nu-\partial_\nu A_\mu-[A_\mu,A_\nu]=F_{\mu\nu},
\nonumber\\
&&\partial_\mu F^{\mu\nu}-[A_\mu,F^{\mu\nu}] =\phi^\dagger \beta i
h^\nu\phi,
\niklabel{DYMYM1}\\
&&\partial_\mu B_\nu-\partial_\nu B_\mu-[B_\mu,B_\nu]=G_{\mu\nu},
\nonumber\\
&&\partial_\mu G^{\mu\nu}-[B_\mu,G^{\mu\nu}] =\frac{3}{16}m^3 i
h^\nu. \nonumber
\end{eqnarray}
This system of equations contains two pairs of Yang--Mills
equations for the fields $(A_\mu,F_{\mu\nu})$ and
$(B_\mu,G_{\mu\nu})$ respectively.

Consider the system of equations (\ref{DYMYM1}), where the
idempotent $t$, the real constant $m$, and the generators of
Clifford algebra $e^a$ are known and the variables
$h^\mu,\phi,A_\mu,F_{\mu\nu},B_\mu,G_{\mu\nu}$ are unknown and
such that
\begin{description}
\item[1a.] The vector $i h^\mu=i h^\mu(x)\in\sp(\cl(1,3))\T^1$
satisfies conditions (\ref{h:cond3}). \item[2a.] The element
$\phi=\phi(x)\in I(t)$ is a scalar of the Minkowski space.
\item[3a.] $A_\mu=A_\mu(x)\in L(t)\T_1$. \item[4a.]
$F_{\mu\nu}=F_{\mu\nu}(x)\in L(t)\T_{[2]}$. \item[5a.]
$B_\mu=B_\mu(x)\in\sp(\cl(1,3))\T_1$. \item[6a.]
$G_{\mu\nu}=G_{\mu\nu}(x)\in\sp(\cl(1,3))\T_{[2]}.$
\end{description}
This system of equation is called  {\em the model
Dirac--Yang--Mills system of equations with two Yang--Mills
fields}.

Suppose that the variables
$$
\phi,\,h^\mu,\,A_\mu,\,B_\mu,\, F_{\mu\nu},\,G_{\mu\nu}
$$
satisfy the conditions {\bf 1a-6a} and satisfy equations
(\ref{DYMYM1}). We see the vector $h^\mu$ at the right hand part
of the Yang--Mills equations
\begin{eqnarray*}
&&\partial_\mu B_\nu-\partial_\nu
B_\mu-[B_\mu,B_\nu]=G_{\mu\nu},\\
&&\partial_\mu G^{\mu\nu}-[B_\mu,G^{\mu\nu}] =\frac{3}{16}m^3 i
h^\nu.
\end{eqnarray*}
Therefore the vector field $h^\mu$ satisfies the non-abelian
conservation law
\begin{equation}
\partial_\mu h^\mu-[B_\mu,h^\mu]=0.
\niklabel{nonabel:cons:new}
\end{equation}
However the identities (\ref{hB:new})  can't be fulfilled. Hence
we may consider the system of equations (\ref{DYMYM1}) as a
generalization of the system of equations (\ref{MDYMW1}).

\medskip

\noindent{\bf Properties of the model Dirac--Yang--Mills
equations}. A transformation of variables in the system of
equation (\ref{DYMYM1}) is called {\em equivalent transformation}
if this system of equation written for transformed variables has
the same form as the system of equation in initial variables. In
this case we say that system (\ref{DYMYM1}) is {\em covariant}
w.r.t. this transformation of variables.

An equivalent transformation of variables in the system of
equation (\ref{DYMYM1}) is called {\em symmetry} if the generators
$e^a$ and the Hermitian idempotent $t$ do not transform (see
\cite{TMP,mybook} for details).

Let us discuss the properties of the model equations
(\ref{DYMYM1}) that related to equivalent transformations and
symmetries. Let
$\Theta=\{h^\mu,\phi,B_\mu,G_{\mu\nu},A_\mu,F_{\mu\nu}\}$ satisfy
the equations (\ref{DYMYM1}).

\num{1.} (Symmetry). All the variables in the system of equation
(\ref{DYMYM1}) are tensors (scalars are rank $0$ tensors).
Therefore this system of equations is covariant under Lorentzian
changes of coordinates.

\num{2.} Consider bilinear forms of the model Dirac--Yang--Mills
equations (\ref{DYMYM1})
$$
i J^{\mu_1\ldots\mu_k}= i^{\frac{k(k-1)}{2}+1}\phi^\dagger \beta
h^{[\mu_1}\ldots h^{\mu_k]}\phi \in L(t)\T^{[k]}.
$$
Bilinear forms $J^{\mu_1\ldots\mu_k}$ are the components of
contravariant antisymmetric tensors of rank $k$ with values in
Hermitian elements of the Clifford Algebra $\cl(1,3)$. Eigenvalues
of these bilinear forms are real.

\num{3.} The vector
$$
i J^\mu=\phi^\dagger \beta i h^\mu\phi\in L(t)\T^1
$$
satisfy non-abelian conservative law
$$
\partial_\mu J^\mu-[A_\mu,J^\mu]=0.
$$

\num{4.} The equations (\ref{DYMYM1}) are covariant under the
following global transformation defined by a unitary element
$U\in\cl(1,3)$, $U^\dagger=U^{-1}$, $\partial_\mu U=0$,
\begin{equation}
\phi\to\phi U, \quad A_\mu\to U^{-1}A_\mu U,\quad F_{\mu\nu}\to
U^{-1}F_{\mu\nu}U,\quad t\to U^{-1}t U.
\niklabel{unitary:glob:trans}
\end{equation}

\num{5.} (Symmetry). The equations (\ref{DYMYM1}) are covariant
under the local (gauge) transformation with $U=U(x)\in G(t)$
\begin{equation}
\phi\to\phi U,\quad A_\mu\to U^{-1}A_\mu U-U^{-1}\partial_\mu
U,\quad F_{\mu\nu}\to U^{-1}F_{\mu\nu}U.
\niklabel{unitary:loc:trans}
\end{equation}

Note that for  $U\in G(t)$ we have $[U,t]=0$ and under the
considered transformation the Hermitian idempotent $t$ does not
transform.

\num{6.} (Symmetry). System of equation (\ref{DYMYM1}) is
covariant w.r.t. the local (gauge) transformation of variables
$\Theta\to\hat\Theta$ induced by an element
$W=W(x)\in\Sp(\cl(1,3))$:
$$
\hat\phi=W^{-1}\phi,\quad \hat h^\mu=W^{-1}h^\mu W,\quad \hat
B_\mu=W^{-1}B_\mu W-W^{-1}\partial_\mu W.
$$

\num{7.} System of equation (\ref{DYMYM1}) is covariant w.r.t. the
discreet transformation (complex conjugation) of variables
$$
i h^\mu\to\overline{i h^\mu},\quad (h^\mu\to-\bar h^\mu), \quad
 t\to\bar t,
$$
$$
\phi\to\bar\phi,\quad A_\mu\to\bar A_\mu,\quad F_{\mu\nu}\to\bar
F_{\mu\nu},\quad B_\mu\to\bar B_\mu,
$$
here we suppose that $\bar e^{a_1\ldots a_k}=e^{a_1\ldots a_k}$.

\num{8.} (Symmetry). System of equation (\ref{DYMYM1}) is
covariant w.r.t. the discreet transformation of variables
\begin{eqnarray}
&&i h^\mu\to\overline{i h^\mu},\quad (h^\mu\to-\bar h^\mu), \quad
\phi\to\bar\phi J,\quad B_\mu\to\bar B_\mu,\nonumber\\
&&A_\mu\to J^{-1}\bar A_\mu J,\quad F_{\mu\nu}\to J^{-1}\bar
F_{\mu\nu}J, \nonumber
\end{eqnarray}
where $J=-e^1e^3$.

\medskip

\noindent{\bf Discussion of the model}. We have introduced system
of equation (\ref{DYMYM1}), which consists of three parts
\begin{itemize}
\item The model Dirac equation
$$
i h^\mu(\partial_\mu\phi+\phi A_\mu-B_\mu\phi)=0
$$
for the wave function $(ih^\mu,\phi)$ of spin $1/2$ particle.

\item The first pair of Yang--Mills equations
\begin{eqnarray}
&&\partial_\mu A_\nu-\partial_\nu A_\mu-[A_\mu,A_\nu]=F_{\mu\nu},
\nonumber\\
&&\partial_\mu F^{\mu\nu}-[A_\mu,F^{\mu\nu}] =\phi^\dagger \beta i
h^\nu\phi,\label{YM:U}
\end{eqnarray}
describes the Yang--Mills field $(A_\mu,F_{\mu\nu})$ with the
gauge group $G(t)$ that is isomorphic to a subgroup of the unitary
group $\U(4)$. According to the Standard Model, if the gauge group
$G(t)$ is isomorphic to one of three groups -- $\U(1)$,
$\U(1)\times \SU(2)$, $\SU(3)$, then system of equation
(\ref{YM:U}) can be used for the description of the
electromagnetic (QED) interaction, the electroweak (EW)
interaction, and the strong (QCD) interaction respectively.

\item The second pair of Yang--Mills equations
\begin{eqnarray}
&&\partial_\mu B_\nu-\partial_\nu B_\mu-[B_\mu,B_\nu]=G_{\mu\nu},
\nonumber\\
&&\partial_\mu G^{\mu\nu}-[B_\mu,G^{\mu\nu}] =\frac{3}{16}m^3 i
h^\nu \niklabel{YM:Sp}
\end{eqnarray}
describes the Yang--Mills field $(B_\mu,G_{\mu\nu})$ with the
symplectic group $\Sp(\cl(1,3))$ of gauge symmetry. The dimension
of the Lie algebra $\sp(\cl(1,3))$ is equal to $10$. Hence the
Yang--Mills field $(B_\mu,G_{\mu\nu})$ describes $10$ types of
spin 1 elementary particles (mediators), which interact with the
initial spin $1/2$ particle (wave function $(i h^\mu,\phi)$). The
model Dirac equation does not contain the mass $m$ of spin $1/2$
particle. We see mass $m$ only at the right hand part of
Yang--Mills equations (\ref{YM:Sp}) in the term $3m^3/16$.
Therefore the constant $3m^3/16$ can be considered as a charge of
spin $1/2$ particle relevant to the gauge field
$(B_\mu,G_{\mu\nu})$.
\end{itemize}

\medskip

\noindent{\bf Conclusion}. In considered model, which based on
equations (\ref{DYMYM1}), spin $1/2$ particles acquire masses by
interaction between these fermions and the (spin 1) gauge field
with symplectic symmetry.
\medskip

\noindent{\bf Acknowledgments}. The paper is partially supported
by the grant of the President of Russian Federation for the
support of scientific schools NSh-7675.2010.1 and by the Program
of the Department of Mathematics of the Russian Academy of Science
``Modern problems of theoretical mathematics''.


\end{document}